\documentclass[preprint]{aastex}

\newcommand{\be}{\begin{equation}}
\newcommand{\ee}{\end{equation}}

\begin{document}
\title{Mining the Metal-Rich Stars for Planets} 

\bigskip
\author{Gregory Laughlin$^1$} 

\bigskip
\affil{$^1$NASA Ames Research Center, Moffett Field, CA 94035}

\begin{abstract} 

We examine the correlation between stellar metallicity and the presence
of short-period planets. It appears that approximately 1 \% of dwarf stars
in the solar neighborhood harbor short-period planets characterized by
near-circular orbits and orbital periods $P<20$ days. However, among the
most metal-rich stars (defined as having $[{\rm Fe/H}]>0.2$ dex), it appears
that the fraction increases to 10\%. Using the Hipparcos database
and the Hauck \& Mermilliod (1998) compilation of
Str\"omgren $uvby$ photometry, we identify a sample of
206 metal-rich stars of spectral type K, G and F which have an 
enhanced probability of harboring short-period planets. Many of these
stars would be excellent candidates for addition to radial velocity
surveys. We have searched the Hipparcos epoch photometry for transiting
planets within our 206 star catalog. We find that the quality of the Hipparcos
data is not high enough to permit unambiguous transit detections.
It is, however, possible
to identify candidate transit periods. We then
discuss various ramifications of the stellar metallicity -- planet
connection. First, we show that there is preliminary evidence for 
increasing metallicity with increasing stellar mass among known planet-bearing
stars. This trend can be explained by a scenario in which planet-bearing
stars accrete an average of 30 $M_{\oplus}$ of rocky material after the 
gaseous protoplanetary disk phase has ended. We present dynamical
calculations which suggest that a survey of metallicities of spectroscopic
binary stars can be used to understand the root cause of the stellar
metallicity -- planet connection.

\end{abstract}

\keywords{stars: planetary systems --- 
stars: metallicity} 

\section{Introduction} 

The discovery of extrasolar planetary systems (Mayor \& Queloz 1995,
Marcy \& Butler 1998, Marcy et al. 2000)
has sparked a great deal of excitement. Nearly
fifty planets have now been detected around other stars, and the total
census has grown to the point where statistical trends
are beginning to emerge. 

There appears, for example, to be a dearth
of objects in the mass range between 10 and 80 Jupiter masses, indicating
that brown dwarfs and planets represent two distinct populations, 
presumably formed via different physical processes. 
There also appear to be at least two different classes of extrasolar planets.
When the orbital period of a planet is less than approximately 20 days, the
orbit tends to be nearly circular, whereas extrasolar planets with longer 
periods are generally on much more eccentric orbits. In this paper, we
will refer to planets with $P<20$ days as ``short-period''.
Tidal circularization is inadequate to explain the circular orbits of 
short-period planets with semi-major axes greater than $\sim 0.1$ AU
(Rasio \& Ford 1996, Hut 1981). It is thus likely that
many of these planets migrated gradually inward
due to interactions with a gaseous protostellar disk (Lin, Bodenheimer \&
Richardson 1995), although {\it in situ} formation is also possible
(Bodenheimer, Hubickyj \& Lissauer 2000). The large eccentricities of
the planets with periods longer than 20 days can arise through
dynamical interactions between multiple planets (e.g. Levison, Lissauer \&
Duncan 1998), perturbations from a bound stellar companion (Holman, Touma
\& Tremaine 1997), and interactions with binary stars in the birth aggregate
(Laughlin \& Adams 1998).

Our own Solar 
System presents two more categories: (1) the terrestrial planets, and
(2) the Jovian planets on long-period, nearly circular orbits. As the
time baseline of the ongoing radial velocity surveys increases, long-period
Jupiter-mass planets in other systems will be detectable (Vogt et al. 2000),
whereas
space-based transit surveys such as the proposed Kepler mission (Borucki et al. 
1997; Koch et al. 1998) will be able to detect extrasolar terrestrial
planets.

In this paper, we focus attention on another emerging trend. 
The stars known
to harbor extrasolar planets tend to be considerably more metal-rich
than the average population I 
star in the galactic neighborhood. Indications of
this correlation were apparent even with the first detections
(Gonzalez 1997, 1998, 1999). 51 Peg,
for example, has a metallicity [{\rm Fe/H}]=0.21 dex (Gonzalez 2000),
which makes it more metal-rich than 98\% of the G dwarfs
in the Solar Neighborhood
(Rocha-Pinto \& Maciel 1996). Other planet bearing stars such as
BD-10 3166, 55 Cnc, and 14 Her, all with $[{\rm Fe/H}]\sim$0.5 dex, 
are among the most metal-rich stars currently known (Castro et al. 1997,
Gonzalez et al. 1999).

Extrasolar planet searches have concentrated effort towards surveys of
solar-type main sequence stars of spectral type ranging from roughly
K5 through F7 (Marcy et al. 2000).
These stars are relatively common, and are also relatively
bright. Furthermore, solar analogs tend to be photometrically stable,
and present a useful variety of absorption lines for radial velocity
determination. The current census of planets is therefore concentrated 
among stars of approximately solar mass. There is, however, no apparent bias 
against metal-poor stars in the current surveys. 
Indeed, at a given mass, metal-poor stars are brighter than metal-rich
stars. For instance, doubling the metallicity of a 1 $M_{\odot}$ ZAMS
star from [Fe/H]=0.00 to [Fe/H]=0.30 leads to a 320 K decrease in 
surface temperature and a factor of 1.3 decrease in luminosity (Forestini 1994).
The current radial velocity searches are essentially magnitude-limited
surveys. Metal-poor stars will thus tend to be overrepresented in the
target lists.

Furthermore, Brown et al. (2000) have examined the detailed time-series
photometry of 34091 stars in the globular cluster 47 Tucanae. Using the
Hubble Space Telescope, they collected 8.3 days of data, and expected to
find roughly 30 transiting short-period planets, based on the occurence
rate among stars in the radial velocity surveys. No planets were found.
The metallicity of 47 Tucanae is [Fe/H]=-0.7, 
which is considerably lower 
than the metallicity of any star known to harbor an extrasolar planet. 

There are now enough planets for us to begin to quantify the short-period
planet -- stellar metallicity correlation, and to make
reasonable estimates of the overall frequency
of occurence of short-period planets.
Rocha-Pinto \& Maciel (1996) have used the Third Catalog of Nearby
Stars (Gliese \& Jahreiss 1991) to identify 345 dwarf stars within
25 pc of probable spectral type G0--G9. Within this volume-limited
(and essentially complete) sample, three stars (51 Peg,
HD 217107, and 55 Cnc) are known to have short-period planets, implying
a frequency of occurence of roughly 1\% among solar type stars. This
estimate suffers from small number statistics, but is in rough agreement
with the initial estimate of 2\% by Butler et al. (1997) based on the number
of short-period planets found in the original Lick radial velocity
survey.

Butler et al. (2000) report that most single 
G, K, and late F dwarf stars of magnitude $V<8$
($\sim 2000$ stars) are presently being monitored by the radial velocity
surveys. The Rocha-Pinto
\& Maciel (1996) census indicates that there should be approximately
1,100 G dwarfs lying within a radius of 40 pc. A majority of these will be
brighter than V=8, and hence should already be under surveillance.
Short-period planets of mass $M_{p} > 0.3 M_{\rm JUP}$ are readily
detected after a year or so of observation. One can thus conclude that 
the six short-period planets orbiting G dwarfs within 40 parsecs should
constitute a reasonable fraction of those which actually exist. It would
thus appear that {\it at least} 0.5\% of solar type dwarfs are accompanied
by short-period planets.

Finally, Vogt et al. (2000) report that among
1000 stars with more than two years of accumulated data, nine planets with
$P<20$ days have been discovered. All of these estimates are in reasonable
concordance, and together
suggest that the rate of occurence of short-period planets among single main
sequence stars in the solar neighborhood is of order 1\%.

As mentioned above, the stars 
which bear planets are (on average) enriched in metals in comparison
to the Sun, and this trend appears to be particularly strong for the
short-period planets (Gonzalez 1997, Butler et al.
2000). In Table 1, we list the properties of currently known planets
which have orbital periods of 20 days or less.
The logarithmic average metallicity of the primary stars harboring
short-period planets is $[{\rm Fe/H}]=$0.16 dex (excluding HD 83443 for which
no metallicity determination or $uvby$ photometry currently
exists), whereas the logarithmic average 
metallicity of G dwarfs in the solar neighborhood is $[{\rm Fe/H}]=$-0.2 dex
(Rocha-Pinto \& Maciel 1996). Of the six G dwarfs in Table 1 which lie
within 40 pc, four have metallicity $[{\rm Fe/H}]>$0.2 dex. Stars of 
metallicity $[{\rm Fe/H}]>$0.2 dex constitute a mere 2.4\% of the local G 
dwarf population, and hence about 
35 such metal-rich G dwarfs are expected to lie within 40 parsecs. It therefore
appears that approximately 10\% of stars with $[{\rm Fe/H}]>$0.2 dex are 
accompanied by short-period planets. In other words, the most metal-rich
stars have a ten times increased probability of harboring a short-period
planet.
In this paper, we seek to explore the
consequences of this remarkable correlation.

This paper is organized as follows: in \S 2 we compile a catalog of 206 
high-metallicity
stars which have an increased likelihood of harboring short-period
planets. In \S 3 we search the Hipparcos epoch photometry for 
candidate transit events among the 206 stars in the catalog, and discuss
the possibility of detecting transiting planets using this method. In \S 4
we discuss an emerging trend toward increasing metallicity as a function of 
stellar mass among the known planet-bearing stars. In \S 5
investigate the possibility of using metallicity measurements of
spectroscopic binary stars to determine whether the short-period planet --
stellar metallicity
correlation arises from the intrinsic properties of the parent star-forming
molecular cloud core or, alternately, through post-formation enrichment.
We summarize and discuss our results in \S 6.

\section{A Target List of High-Metallicity Stars} 

The clear existence of a correlation between high stellar metallicity and
the presence of a short-period planet indicates that a selective targeting
of high metallicity stars would constitute an effective search method. Indeed, 
this strategy has already yielded success. The formerly obscure star BD-10 3166 
was listed by Castro et al. (1997) as one of the five most
metal-rich stars known, with $[{\rm Fe/H}]=0.5$.
As reported in Butler et al. (2000), it was added 
to the Keck Radial Velocity Survey at the suggestion of Gonzalez, Wallerstein
\& Saar (1999), and was soon thereafter found to harbor a $M \sin i = 0.48
M_{\rm JUP}$ planet in a 3.487 day low-eccentricity orbit.

Although most dwarf stars with $V<8$
are currently being targeted by the various radial 
velocity
searches, the total number of stars being surveyed is much 
smaller
than the total population of dwarf stars now known to exist within the broader 
($d<100$ pc) solar neighborhood.
It is therefore of interest to compile a list of metal-rich stars
that are dim enough to have likely eluded detailed attention thus far.

Spectroscopic abundance analyses exist for only a relatively small number
of stars, but there are over 63,300 stars for which $uvby$ photometry
has been reported in the literature. Furthermore, every  individual $uvby$ 
measurement
published prior to the middle of 1996 has been compiled in machine-readable
form by Hauck \& Mermilliod (1998).

Schuster \& Nissen (1989)
have published a calibration which uses the standard $uvby$ indices,
$b-y$, $m_{1}$, and $c_{1}$  to estimate the metallicity of a given star:
$$[{\rm Fe/H}]=1.052 - 73.21 \, m_{1} + 280.9 \, m_{1} (b-y) +
333.95 \, m_{1}^{2}(b-y) - 595.5 \, m_{1} (b-y)^{2} +$$
$$(5.486 - 41.62 \, m_{1} - 7.963 \, (b-y)) \, \log [m_{1} 
-(0.6322 - 3.58 \, (b-y) + 5.20 \, (b-y)^{2})]$$
for $0.22 \le (b-y) < 0.375$, and:
$$[{\rm Fe/H}]=-2.0965 + 22.45 \, m_{1} - 53.8 \, m_{1}^{2} -
62.04 \, m_{1} (b-y) + 145.5 \, m_{1}^{2}(b-y) +$$
$$(85.1 \, m_{1} -13.8 \, c_{1} - 137.2 \, m_{1}^{2}) c_{1} \eqno{(1)}$$
for $0.375 \le (b-y) \le 0.59$. 
The average uncertainty for
this calibration is stated to be $\Delta [{\rm Fe/H}] = 0.16$ dex.

This calibration is applicable to the dwarf stars of spectral
type ranging from F through early K which have been preferentially
targeted in the radial velocity surveys.
Rocha-Pinto \& Maciel (1996)
applied the Schuster \& Nissen (1989) calibration to a sample of 79 G dwarfs
in the solar neighborhood for which spectroscopic abundances have
been determined, and found very good agreement across a wide
range of metallicities. In particular,
Rocha-Pinto \& Maciel (1996) found that
the Schuster \& Nissen (1989) calibration shows better agreement with
the spectroscopic abundances than does the calibration of Olsen (1984).
In Figure 1, we compare $uvby$ metallicities 
obtained by applying Equation 1 to the aggregate of planet-bearing stars 
for which alternate, mostly spectroscopic, determinations of the
metallicities are available
(see Butler et al. 2000 for references to individual determinations).
For this high-metallicity subsample, the Schuster \& Nissen (1989)
calibration tends to {\it underestimate} the metallicity of a given
star by an average of 0.06 dex
in comparison to the spectroscopically determined value.
We can thus conclude that the $uvby$
calibration is sufficiently accurate for use in selecting a statistically
metal-rich sample of stars from the Hauck \& Mermilliod (1998) compilation.

The catalog shown in Table 2 was obtained with the following procedure.
The database of Hipparcos target stars (Perryman et al. 1997)
was queried for objects with (1)
magnitudes in the range $7.8 < V < 11$, (2) parallaxes $\pi>0.01$ arcseconds,
and (3) color indices in the range
$0.2 < B-V < 0.9$. This query yields a subset of 10,101 stars, most of 
which have HD identifiers.
Within this subset, a second search was made
for stars listed in the Hipparcos catalog as non-multiple and non-variable
and which also appear in the Hauck
\& Mermilliod (1998) compilation. This cut yielded 4825 stars. 
[{\rm Fe/H}] estimates were then computed from each star's {\it uvby} 
photometry. We find that 206 stars (4.27\% of the sample) have [{\rm Fe/H}]
values in excess of the Hyades metallicity of 0.125. 
Given the tendency of the Schuster \& Nissen (1989) calibration to 
underestimate metallicity, these stars are
quite likely to be very metal-rich, and, as such, should have an
approximately tenfold increased
probability (in comparison to field stars) of bearing short-period planets.

There are 31 stars in Table 2 with $V$ magnitudes in the range $7.8 < V <8.0$.
Hence, as stated in Butler et al. (2000), they are thus likely to
already be under
radial velocity surveillance. Indeed, out of these 31 bright 
stars, eight
are currently included in the Keck survey (Marcy, 2000 personal communication).
Of these eight stars, one, (HD 187123) has a planet with $M \sin i=0.54$
$M_{\rm JUP}$ and a period of 3.097 days (Butler et al. 1998). 
One planet in eight stars is consistent with our expected occurence
rate of 10\%.

We recommend that the stars in Table 2 be considered for targeting in
planetary searches. The stars which are later than spectral type F5--F7
can be added to the radial velocity surveys currently underway, and all of the 
stars can be photometrically monitored for transits. In the next section,
we describe a preliminary transit search of our catalog which makes use of
the Hipparcos epoch photometry.

\section{A Preliminary Search for Transiting Planets Among the 
High-Metallicity Catalog Stars}

Given the frequency estimates in the introductory section, 
we expect that roughly 10\%
of the stars listed in Table 2 harbor short-period planets. 
Geometric arguments indicate that approximately
10\% of short-period planets will display transits (Henry et al. 2000),
so it is likely that
there are 1--2 transiting planets orbiting stars in Table 2.

The roughly 10\% fraction of short-period planets which exhibit transits are 
extremely useful. The orbital inclination of a transiting planet is determined
to high accuracy, breaking the $\sin i$ degeneracy which plagues the
radial velocity method of detection, and giving an 
unambiguous mass determination. A transit detection also yields a size
and density for the planet. As more transiting planets are found,
their distribution in the space delineated by mass, radius, and temperature
will be instrumental in allowing for comparison with theoretical models.
Each new transit will help to further
unravel the structure and evolution of the extrasolar planets.

At present, only one transiting extrasolar planet is known. This object
(HD 209458b) orbits its parent star with a period of 3.525 days,
has a mass of $0.62 M_{\rm JUP}$, and a radius of $1.42 R_{\rm JUP}$.
The parent star HD 209458 is a G0 main sequence dwarf with no known 
stellar-mass companions. Based on
Hipparcos measurements of $V=7.65$, $B-V=0.594$, and $d=47$ pc,
Allende Prieto \& Lambert (1999) estimate a mass of 1.03 $M_{\odot}$, and a 
radius of 1.15 $R_{\odot}$ for the star.
HD 209458b was initially discovered with the radial velocity method
(Henry et al. 2000, Mazeh et al. 2000). Subsequent measurements of
the photometric light curve of the star revealed that the planet
was transiting, with a transit duration of 3.1 hours, and a maximum transit
depth of 1.6\% (Henry et al. 2000, Charbonneau et al. 2000).

The primary mission of the Hipparcos satellite was to produce astrometric
information. However, in addition to positional astronomy, the satellite
also obtained an extensive database of photometric measurements during
its 3.36 year mission. This Hipparcos epoch photometry (Perryman et al. 1997,
van Leeuwen et al. 1997)
covers 118,204 stars and contains an average of $n$=110 observations per star.
Almost immediately
after the announcement of the HD 209458 transit detection (Henry et al. 2000,
Charbonneau et al. 2000) several
authors demonstrated that the transit is visible in the Hipparcos epoch
photometry (S\"oderjhelm 1999, Castellano et al. 2000, Robichon \& Arenou
2000). 

Figure 2 shows the 89 photometric measurements which the Hipparcos
satellite made of HD 209458 (HIP 108859), folded at a period of 3.5247
days (consistent with the period obtained from the radial velocity
observations).
The error bars attached to the data points correspond to the quoted 
uncertainty of each individual measurement. The heavy solid line shows
an approximation to the transit lightcurve in which the assumed transit
depth is 1.6\%, and the transit duration is 3 hours (see below).
Note that the slopes of ingress and egress have been omitted.
The transit is manifested as a rather unremarkable group of six
measurements clustered around orbital phase 0.7. Castellano et al. (2000),
however, have shown that with an {\it a priori} knowledge of the
approximate period, such a cluster of low points has approximately
one chance in $5 \times 10^{4}$ of occuring by chance. Robichon \& Arenou
(2000) find a similar result, and suggest that additional transiting 
planets might be detectable within the Hipparcos epoch photometry.
It is thus informative to carry out a preliminary search for the 1--2
transiting short-period planets that are expected to exist among the
high metallicity stars in Table 2. This is especially true for stars of
spectral type F6 and earlier which are inaccesible to radial velocity
surveys (due to a lack of adequate spectral lines for velocity 
cross-correlations).

An automated routine was written to search for transiting planets in the
Hipparcos photometry. For a particular star, the stellar
mass and radius are estimated from the spectral type (see Allen, 2000).
A hypothetical transiting planet is assumed to have a radius of 1.3 
$R_{\rm JUP}$. This value is consistent with the observed radius of
HD 209458b, and it agrees with theoretical expectations for the sizes
of the so-called ``Hot Jupiters''. In a study of the 51 Peg system,
Guillot et al. (1996) used a detailed non-adiabatic structure model 
to show that the radii of 8 Gyr old planets in four day orbits
exhibit a very narrow range of sizes varying from 1.2 $R_{\rm JUP}$
at 0.5 $M_{\rm JUP}$ to 1.4 $R_{\rm JUP}$ at 3.0 $M_{\rm JUP}$.
Given the stellar properties and the radial size of the hypothetical transiting
planet, we next choose a trial orbital period $P$. Transits are assumed
to occur along the major chord of the star as seen from Earth.
The orbital period is therefore sufficient
to fix the transit duration, and the relative sizes of planet and star
fix the transit depth,
yielding an approximate light curve $y_{p}(\phi)$. The
effects of limb darkening and transit ingress and egress are not considered. 
The large intrinsic scatter and the sparse sampling of the Hipparcos photometry
obviates any benefit from these refinements.

For each star in Table 2, the Hipparcos photometry annex lists the
observation epoch (referred to Julian day 2,440,000.0),
the Hipparcos magnitude $H_{p}(i)$, the magnitude
uncertainty $\sigma_{H}(i)$, and a data quality flag $f(i)$, for each 
photometric measurement (van Leeuwen et al. 1997).
We reject data points for which 
$f(i)>8$. The remaining $n$ points are phased modulo the trial orbital
period $P$. A chi-squared statistic
$$\chi^{2}(P,\phi_{j})= {\sum_{i=1}^{n}} ({H_{p}(i)-y_{p}(i,\phi_{j}) 
\over {\sigma_{H}(i)}})^{2} \eqno{(2)}$$
is computed succesively for 500 equally spaced trial transit phases $\phi_{j}$
spanning the range from 0 to 1, and the minimum chi-squared value is
noted. 

The procedure is then applied to 90,000 trial periods in the 1-10 day range,
with $\Delta P=$0.0001 days. Figure 3 shows $\chi^{2}_{min}(P)$ for HD 209458,
for periods in the subinterval ranging from three to four days. The correct 
transit period
of 3.5247 days yields $\chi^{2}_{min}(3.5247)$=115.57, for a transit centered 
at phase $\phi$=0.704. This value for $\chi^{2}$ is the second-best fit to
the data within the 3--4 day range. A spurious period of 3.7909 days yields
a slightly lower value of $\chi^{2}_{min}(3.7909)$=115.50, and there
is a third fit at 3.2952 days which is nearly as good. 

The HD 209458 measurements (which are by no means atypical within the epoch
photometry annex) indicate that unambiguous transit detections are not possible
with typical Hipparcos data. Nevertheless, the quality of the HD 209458
detection is high enough to warrant a survey of possible candidate events 
(period determinations) within 
the high metallicity catalog. The above-described procedure was applied
to all 206 stars in Table 2, and periods with low chi-squared values
were noted. These good-fit periods were then screened according to the
several criteria chosen to enhance the likelihood of a period fit being
a true detection. First, periods were required to fall in the range
$3 < P < 4$ days. This period range is seemingly preferred by actual
systems; as mentioned above, nearly half of the known short period planets
fall in this one day interval. Second, only events with particularly low 
minima were chosen. Candidates are required to have unit weight error
$u = \sqrt{\chi^{2}/{n}} < 1$. 
We also require that a candidate transit interval contain
no more than 2.5 times the expected number of photometric measurements.
In practice, particularly low chi-squared fits
are often achieved by folding a large number of outlying points into a
spurious
transit interval. Nevertheless, true detections are enhanced by a lucky
distribution of points. In the case of HD209485b, 
the six transiting points which contribute to
the 3.5247 day period shown in Figure 3 are 2.25 times the number expected among
89 randomly distributed measurements. (Note that the Hipparcos photometric
measurements are not distributed randomly in time. In many cases,
successive measurements are either spaced very closely or interspersed with
longer gaps than would be expected from a random distribution.) 
Candidate transits which meet these criteria are listed in Table 2. 
For these stars, we list (1) $n$, the number of acceptable photometric
measurements for the star, (2) $P$, the best-fit period, (3) $u$, the
unit weight error of the best chi-squared fit, and (4) $n_{t}/n_{expected}$,
the ratio of observed to expected points in the transit interval.
If multiple candidate periods are found for a particular star, only the
period with the lowest chi-squared fit is listed.
An example candidate transit detection is shown in Figure 4.
The interested reader can examine other candidate transits in Table 2
by entering the Hipparcos identifier
and the trial period into the folding algorithm featured at the Hipparcos
website:
http://astro.estec.esa.nl/SA-general/Projects/Hipparcos/apps/PlotCurve.html.

\section{A Connection Between Metallicity and Stellar Mass for Planet-Bearing Stars}

There appear to be two broad classes of explanation for the correlation
between short-period planets and high stellar metallicity (see e.g.
Gonzalez 1997). The first explanation posits that high metallicity parent
clouds lead more readily to planet formation, and hence to a higher
rate of occurence of short-period planets. High values for $Z$ mean that
more rocky material is available in the nebula to form protoplanetary cores.
Furthermore, the presence of nebular dust above a particular
threshold may trigger vigorous planet formation. A possible
explanation might proceed as follows:
The paradigm scenario for the formation of planetesimals is that
dust grains settle to the midplane of the disk where they become
gravitationally unstable (Goldreich \& Ward 1973). Within the last
decade, however, it has been shown that turbulence driven by the
vertical shear in the orbiting disk can prevent dust from settling
into the thin layer required to activate gravitational instability
(see Weidenschilling \& Cuzzi 1993). Recently, however, Youdin \& Shu
(2000, private communication) have shown that inclusion of off-diagonal
Brunt-V\"ais\"al\"a frequencies can considerably reduce the effectiveness 
of the vertical shear in stirring up turbulence (see also Sekiya 1998).
They find that if the dust density is more than twice the value
expected in the minimum solar nebula,
then the turbulence becomes ineffective for supressing the
Goldreich-Ward (1973) instability.
Photoevaporation of a protostellar disk is one mechanism which can increase
the dust-to-gas ratio to the required threshold, and a high
metallicity nebula would require a smaller
degree of photoevaporation to trigger instability, thus allowing for
vigorous planet formation within the inner regions of a metal-rich disk.

The second hypothesis for the planet-metallicity correlation
holds that systems which form short-period planets 
(independently of the initial metallicity)
are also able to efficiently enrich their parent stars. 
The large observed metallicities therefore occur
only in the convective envelope, and not within the star
as a whole. This can occur by [1] complete inward migration of 
(metal-rich) planets onto the star in response to
tidal interactions with the protostellar disk (Lin, Bodenheimer
\& Richardson 1996), [2] via the scattering of metal-rich
planets onto the star (Rasio \& Ford 1996), or
[3] through the dumping of debris such as planetesimals, comets or
asteroids onto the star (Murray et al. 1998, Quillen \& Holman 2000).

Main sequence stars which are more massive than $\sim 0.25 M_{\odot}$ and
less massive than $\sim 1.4 M_{\odot}$ have a radiative central core 
surrounded by a convective envelope. The more massive the star, the
less mass is contained in the convective envelope. In the 
present day Sun, for example, the convection zone contains 0.02
$M_{\odot}$ and extends over the outer 26\% of the Sun's radius
(Sackmann, Boothroyd, \& Kraemer 1993). The gas
giant planets in our solar system appear to be enriched in heavy elements
(partially as the result of their rock-ice cores), and are likely to
have metallicities of order $Z \ge 0.1$. The dissolution of Jupiter (assuming
$Z_{\rm JUP}=0.1$) into 
the convective envelope of the present day Sun would thus raise the metallicity
of the convective envelope by 0.08 dex. By contrast, adding Jupiter
to a 0.8 $M_{\odot}$ ZAMS solar metallicity main sequence star, which has
a 0.73 $M_{\odot}$ convective envelope (Forestini 1994), yields a metallicity
increase of only 0.025 dex. Folding Jupiter into the
0.007 $M_{\odot}$ convective envelope of a 1.2 $M_{\odot}$ ZAMS star
gives a large metallicity increase of 0.176 dex.

Figure 5 indicates that there is some evidence for increasing metallicity with
increasing stellar mass among all known planet-bearing stars for which
metallicities are available. Most of the masses and metallicities
shown in this plot were taken from Table 5 of Butler et al. (2000). Many of 
these metallicities result from spectroscopic determinations. In addition,
four of the five new planet-bearing stars announced by Mayor et al. (2000),
and the three new planet-bearing stars announced by Fischer et al. (2000)
are also included. The planet-bearing stars
HD 177830, GJ 876, and HD 83443 are not included
in the figure, since neither high resolution spectroscopy nor $uvby$ 
photometry is yet available for these stars. A linear
regression to the data yields a best fit slope of 0.548 ${\rm dex}/M_{\odot}$.

Stars which are heavier than roughly 1 $M_{\odot}$ have 
main sequence lifetimes which are
shorter than the age of the galactic disk. In general, therefore, a 
population of F dwarfs will be younger than a population of G dwarfs. For
example, assuming a constant rate of star formation in the Galaxy, and
the stellar mass-lifetime relation given by Iben \& Laughlin (1989), the
average age of a sample of 1 $M_{\odot}$ G dwarf stars is 4.15 Gyr, whereas
the average age of an aggregate of 1.3 $M_{\odot}$ F dwarf stars would be only
1.7 Gyr.

Edvardson et al. (1993) examined a sample of nearby F and G stars, and
found that the average metallicity in their sample was a gradually 
decreasing function of age. Ng \& Bertelli (1998) further quantified 
this relation, and found that the age-metallicity relation for the 
averaged population displays a small, but nevertheless discernable slope
of 0.07 dex/Gyr. This result is in accordance with the enrichment
of the galactic disk as a result of ongoing stellar evolution, and it
means that a local sample of F dwarfs should be statistically more 
metal-rich than a similar sample of G dwarfs. Much of the enrichment,
however, is reflected in the paucity of very metal-poor main sequence
F stars.

We can check the degree to which the trend in Figure 5 is due to 
the younger average age of the F stars by a constructing a local, 
volume-limited comparison sample of G, F, and early K dwarfs: We
queried the Hipparcos catalog (Perryman et al. 1997) for all objects
lying within 25 pc, and then removed the stars which are listed
as either multiple (Hipparcos field ${\rm H58}>1$) or variable.
From this volume-limited subset, we retained all stars for which 
$uvby$ measurements exist, and for which  $0.22<b-y<0.59$. The masses
of these stars were then obtained from the database of Prieto \& Lambert (1998),
and metallicities were computed
using the Schuster \& Nissen (1989) calibration.
This yields an essentially complete, volume-limited sample of 196 G and F
dwarfs which have metallicities in excess of [Fe/H]=-0.4, and masses in the 
range 0.88 $M_{\odot}$ -- 1.45 $M_{\odot}$. These stars
are plotted as open circles in Figure 6. For comparison,
the metallicities of the planet-bearing stars were recomputed using the
Schuster \& Nissen (1989) calibration) and plotted as filled circles.
As mentioned in Butler et al. (2000), most of the mass estimates for the 
planet-bearing stars in Figure 5 are from the Prieto \& Lambert (1998)
database, so a comparison of [Fe/H] vs. $M$ for the two populations is
readily justified.

The use of the Schuster \& Nissen (1989) calibration in Figure 6 
maintains a noticeable metallicity-mass gradient among the
planet-bearing stars  of 0.498 ${\rm dex}/M_{\odot}$, yet there is only
a marginal -0.064 ${\rm dex}/M_{\odot}$ gradient among the stars 
in the local sample with $\rm{[Fe/H]}>-0.4$. This nearly null slope
for the local aggregate is in good agreement with the conclusion of
Rocha-Pinto \& Maciel (1998), who find that there is a ``remarkable
consistency amongst the [metallicity] distributions of F, G, and
K type dwarfs.''

Furthermore, Figure 6 shows that for $M<M_{\odot}$, the planet-bearing stars are
drawn essentially randomly from the metallicity distribution of
stars having $\rm{[Fe/H]}>-0.4$. By 1.2 $M_{\odot}$, however, the
planet-bearing stars lie in the region occupied by the most metal-rich
stars of the local distribution. The two planet-bearing stars
with $M \ge 1.4 M_{\odot}$ (HD 38429 and HD 89744) are more metal-rich
than any star of similar mass within 25 pc.

As a general trend, solar-type stars begin their pre-main-sequence
evolution on the Hayashi track with a fully convective stellar configuration.
After several million years, radiative cores appear and the size of the
convective envelope steadily decreases over the next 10 million years or so. 
The lifetimes of protostellar disks are on the order of several million years.
As discussed by Laughlin \& Adams (1997), if planets are added to the
star primarily as a result of inward migration caused by tidal interactions
between the disk and its embedded planets, then a
metallicity gradient as a function of stellar mass is likely to exist only
among early F-type and A-type stars (1.5 --2.5 $M_{\odot}$). Stars
in this mass range have relatively low-mass convective envelopes during the 
phase when a protoplanetary disk is present. Laughlin \& Adams (1997) 
examined the metallicities of a volume-limited sample of F dwarfs, and
found only marginal evidence for such a trend.
If the much more dramatic trend shown in Figure
5 among the K-type through late F-type type planet-bearing stars is real, then 
the enrichment processes must be occuring over timescales which are
considerably longer than the typical disk lifetime. The long-term process of
resonant scattering of planetesimals into the parent star would thus provide 
the most natural mechanism for enrichment.

\section{Using Binary Stars to Probe the Metallicity Correlation}

Short-period binary stars can provide another observational test
to select between the initial metallicity hypothesis and the enrichment
hypotheses. The basic idea is straightforward. A close binary star,
once formed, will tend to protect itself against accretion of either
planets or other debris. Material which wanders within several 
binary separations of the central stars tends to be ejected from 
the system as a result of three-body interactions. 

As a first illustration of this principle, one can imagine a circumbinary
disk which has formed a Jupiter-mass planet. Tidal interactions between the 
disk and the planet are then imagined to cause the planet to migrate inwards
towards the central binary star. This process can be simulated by placing
Jupiter-mass particles in
2 AU circular orbits around the center of mass of a 0.25 AU
zero-eccentricity binary with equal componet masses of 0.5 $M_{\odot}$.
The tidal effect of the circumbinary 
disk is modeled as a tiny, constant azimuthal torque
which causes the planetary orbit to spiral inward on a $10^{5}$ year timescale.
Once the orbit has decayed to the point where the planet reaches an orbital
radius of several binary separations, 3-body interactions tend to eject
the planet from the system. A typical example of this process is illustrated 
in Figure 7.

Using a Bulirsch-Stoer integration scheme, we have computed 400 different
simulations of inward planetary migration onto the binary star system
described above. Variations in the initial conditions were obtained simply
by varying the initial orbital phase of the Jupiter-mass planet in its
2 AU circular orbit. We find that in 356 cases (89\% of the time), the planet
is ejected from the system. We found that 23 of the simulations resulted in the 
planet colliding with the first star, and that 21 of the simulations ended in 
collision with the second star.
As we discussed above, if the mass-metallicity trend shown in Figures 5 and 6
continues to hold as more planets are discovered,
then it is rather unlikely that the planet -- stellar
metallicity correlation results from accretion of planets during the phase 
when a gaseous protostellar disk is present. However, if accreting planets 
really are responsible for the metallicity trend, it is clear that close
binary stars can eject most planets which migrate into their vicinity, and 
one would expect a deficit of super-metal-rich spectroscopic binaries, 
especially among the easily enriched A and F type stars.

Once the gaseous nebula has dispersed, enrichment of the convective envelope
can occur as long-term dynamical instabilities among the planets, 
planetesimals and debris still present in the system cause material to be 
cast onto the fully formed star. Again, however, close binaries should be
more adept than single stars at keeping such material at bay.

To illustrate how binary self-protection works in the longer-term context of
resonant scattering, we have computed
two sets of long-term integrations of a simplified model of a planetary
system. In the first set of integrations, two planets are placed in
orbit around a solar mass star. The inner planet is given a semi-major axis
$a=5 AU$, and an eccentricity $e=0.05$. The planetary mass is assumed to
be $1.0 \times 10^{-3} M_{\odot}$. These parameters are quite close to
the present day properties of Jupiter. The second planet is given a
mass of $6.0 \times 10^{-4} M_{\odot}$, an eccentricity $e=0.1$, and
a semi-major axis $a=9 AU$. These values were chosen in order to yield
a system which is similar to, but dynamically slightly more active than
our own Jupiter-Saturn pair.

256 test particles were placed on circular orbits around the
central star, with initial semi-major axes ranging in equally spaced increments
from 1.0 to 3.6 AU. The system was then integrated forward in time for one
million years using a Bulirsch-Stoer integration scheme (Press et al. 1996).
The eccentricities
and semi-major axes of the test particles were frequently evaluated.
In Figure 8, the maximum eccentricity achieved by each particle is 
indicated by an open circle plotted with respect to the initial semi-major 
axis. 

Figure 8 indicates that many orbital trajectories within the two planet 
model system suffer large increases in eccentricity. The broad resonant feature
centered at approximately 2.1 AU is the analog of the $\nu_{6}$ secular
resonance in the Solar System.
This resonance arises because the induced
precession rates of the perihelia of the test particles in this region 
are commensurate with the precession of the perihelion of the outer
saturn-like planet (see e.g. Murray \& Dermott 1999). Powerful resonances
of this sort are expected to be common when a system contains more than
one giant planet.
Other resonance features are also visible. At 3.15 AU, for example, the 2:1 mean
motion resonance with the inner Jupiter-mass planet leads to a significant
eccentricity increase for the particles having initial semi-major axes that
place them in this region.

Particles which find themselves in strong resonances such as the $\nu_{6}$
rapidly develop very large eccentricities. In many calculations, the 
eccentricitity grows so large that the perihelion,
$r_{p}=a(1-e)$, of the particle is smaller than the stellar radius.
When this situation occurs, the object is destroyed by the star, and its
burden of heavy elements is mixed through the convective envelope
(see Quillen \& Holman 2000).
Alternately, some particles which develop extreme eccentricities also
experience increases in semi-major axis which cause their orbits to cross
the orbits of the planets. In such cases, the particles are often
ejected from the system without encountering the central star. In Figure
8, the particles which crash into the central star are indicated by 
filled black circles.
The wide $\nu_{6}$ resonance is seen to be the most effective mechanism
for delivering debris to the surface of the central star.

Figure 8 also shows the results of integrations in which the same system
of two planets was
started in circumbinary orbit around a pair of 0.5 $M_{\odot}$ stars
on an $e$=0.0 orbit of separation $a=0.25$ AU.
A total of 220 test particles were placed on circular orbits in the
region between 1.5 and 3.8 AU.
In this alternate system,
the strong high-frequency perturbation arising from the binary potential
destroys or weakens many of the resonances. In particular, the $\nu_{6}$
resonance, which efficiently directed material onto a single central star, is
now completely absent.
Furthermore, those particles which do receive large eccentricities
tend to suffer close encounters with the perturbing planets and are ejected
from the system without encountering the central stars, and the small fraction
of particles that do manage to find themselves in the vicinity
of the central stars stand a high probability of being ejected.
Out of the 220 test 
particles in the simulation, none were observed to hit either 
member of the binary pair.

The foregoing calculations are by no means exhaustive or definitive, but
the suggestion is clear. A close binary star is capable of protecting
itself against
enrichment. If enrichment is responsible for the enhanced metallicity
of planet-bearing stars, then there should be a statistical deficit
of super-metal-rich short period binary systems.

Can such a statistical deficit be observed? For spectroscopic binary stars,
individual double lines will represent an added technical difficulty for
spectroscopic determinations of the metallicity. One must separate the
composite spectrum into two spectra corresponding to the individual stars.
There have been some attempts to obtain metallicities for double-lined
eclipsing spectroscopic binaries in order to make the eclipsing systems
a better test of stellar evolution models (e.g. Ryabchikova et al. 1999). 
However, there are not yet enough determinations to constitute a statistically
meaningful sample, and the metallicities which have been obtained are not 
as accurate as
for single stars (G. Gonzalez, personal communication).

If the two binary components are of the same spectral type (as would be
the case in the system used for our test integrations) then $uvby$ photometry
gives a meaningful metallicity estimate if one assumes that the 
metallicities of both stars are the same. Alternately, for systems with
extreme mass ratios -- an F8, say, and an M8 -- where the flux from
the secondary is swamped by the flux from the primary, it should be 
possible to get a determination of the metallicity of the primary
component either with a spectroscopic analysis or with $uvby$ photometry.
In any case, a comprehensive survey of metallicities of close binary
systems would be a very valuable resource.

\section{Discussion} 

We have studied the emerging correlation between high stellar
metallicity and the presence of extrasolar planets. This correlation
is one of several unexpected and exciting results which have stemmed
from the growing census of extrasolar planets. Our results are 
summarized as follows:

1. The frequency of occurence of short-period planets (those with
periods $P<20$ days) is approximately 1\% among solar type stars.
Among the most metal-rich stars ($[{\rm Fe/H}]>$0.2) this fraction increases
to approximately 10\%. 

2. Metallicity estimates using the Schuster \& Nissen (1989) calibration
of $uvby$ photometry
yield [Fe/H] values which are well correlated with metallicities derived
from detailed spectroscpic analysis (see Figure 1). The $uvby$ calibration,
however, systematically underestimates metallicities by $\sim$ 0.06 dex for
stars of solar metallicity and above.

3. Using the Hipparcos database (Perryman et al. 1997), and the Hauck
\& Mermilliod (1998) listing of $uvby$ photometry, we have compiled a
catalog of 206 metal-rich stars which are dim enough ($V>$8) to have
likely eluded inclusion in the current radial velocity surveys (Table 2). We
advocate that these stars be added to ongoing surveys, due to their
high likelihood of possessing short-period planets. It is important
to locate more short-period planets in order to be assured of further
transiting systems. Transits are very important because they
allow planetary properties (mass, radius, density) to be uniquely 
determined. We should mention, however, that the new systems discovered
within this candidate list should not be used in future statistical
analyses of the incidence rate of planets discovered with the Doppler
method.

4. We have used the Hipparcos epoch photometry to
carry out a search for transiting planets associated with
the stars in our high-metallicity catalog. We show that in the absence
of {\it a priori} information regarding the period, the quality of 
the Hipparcos photometry is insufficient to make unambiguous detections.
The data quality is, however, high enough to identify candidate events
(see for example Figure 4),
and we have listed a number of candidate periods associated with the
stars in our catalog.

5. There is a preliminary suggestion of a trend toward increasing
metallicity with increasing stellar mass among the aggregate of known
planet-bearing stars (Figure 5). This trend is consistent with a scenario
in which systems capable of forming short-period planets are also 
capable of enriching the convective envelopes of their parent stars
after the gaseous protoplanetary disks have dispersed.
As shown by Figure 6, the trend does not appear to stem from the systematic
differences in age between F and G stars.

6. We have explored the possibility of using spectroscopic binary
stars to determine the underlying cause of the extrasolar planet --
stellar metallicity correlation. If the large stellar metallicities are
due to a post-formation enrichment process, then super-metal-rich
spectroscopic binaries should be rare in comparison to their single
counterparts.

As more planets are discovered, the true extent of the correlations 
discussed here will become evident. In particular, it will be interesting
to see whether
the correlation continues to hold for systems like our own, in which the
Jovian-mass planets have orbital periods of a decade or longer. Such systems,
if they are common, will soon be uncovered by the radial velocity surveys.

The work in this paper represents a first effort to exploit the
planet -- metallicity connection, and hence there is a great deal of
additional interesting research which could be done. For example,
it would be worthwhile to search the brightest early F dwarfs in the Hipparcos
database for transits. The early F dwarfs are inaccessible to planet detection
via the radial velocity method, and have not yet been monitored
for planets. These stars, however, 
have considerably smaller relative errors in their photometric measurements,
and are hence considerably better suited to transit detection than the 
dim stars analyzed in Table 2. 
The dynamical calculations in this paper would benefit from a wider
survey of parameter space in order to better quantify the ability of 
close binary stars to protect themselves against enrichment. 
Additionally, as discussed in section 5, it will be very useful to
obtain the metallicities of a statistically significant sample of spectroscopic
binary stars.

The existence of a connection between high stellar metallicity and the
presence of a short-period planet is one of the most exciting developments
that has accompanied the discovery of extrasolar planets. This remarkable
correlation is telling us something profound about the formation and
evolution of planetary systems. 

\subsection{Acknowledgements} 

I would like to thank Fred Adams, Douglas Caldwell, Tim Castellano, 
John Chambers, Debra Fischer Guillermo Gonzalez, Geoff Marcy, Doug
Lin, \& Andrew Youdin for
discussions which made this work possible. 
Guillermo Gonzalez provided a helpful referee's report, which led to
the inclusion of Figure 6 in the manuscript.
This research has made use of the SIMBAD database, operated at 
CDS, Strasbourg, France.
This work was supported by a NASA
astrophysics theory program which supports a joint Center for Star
Formation Studies at NASA-Ames Research Center, UC Berkeley and UC
Santa Cruz.

\clearpage

\begin{figure}
\plotone{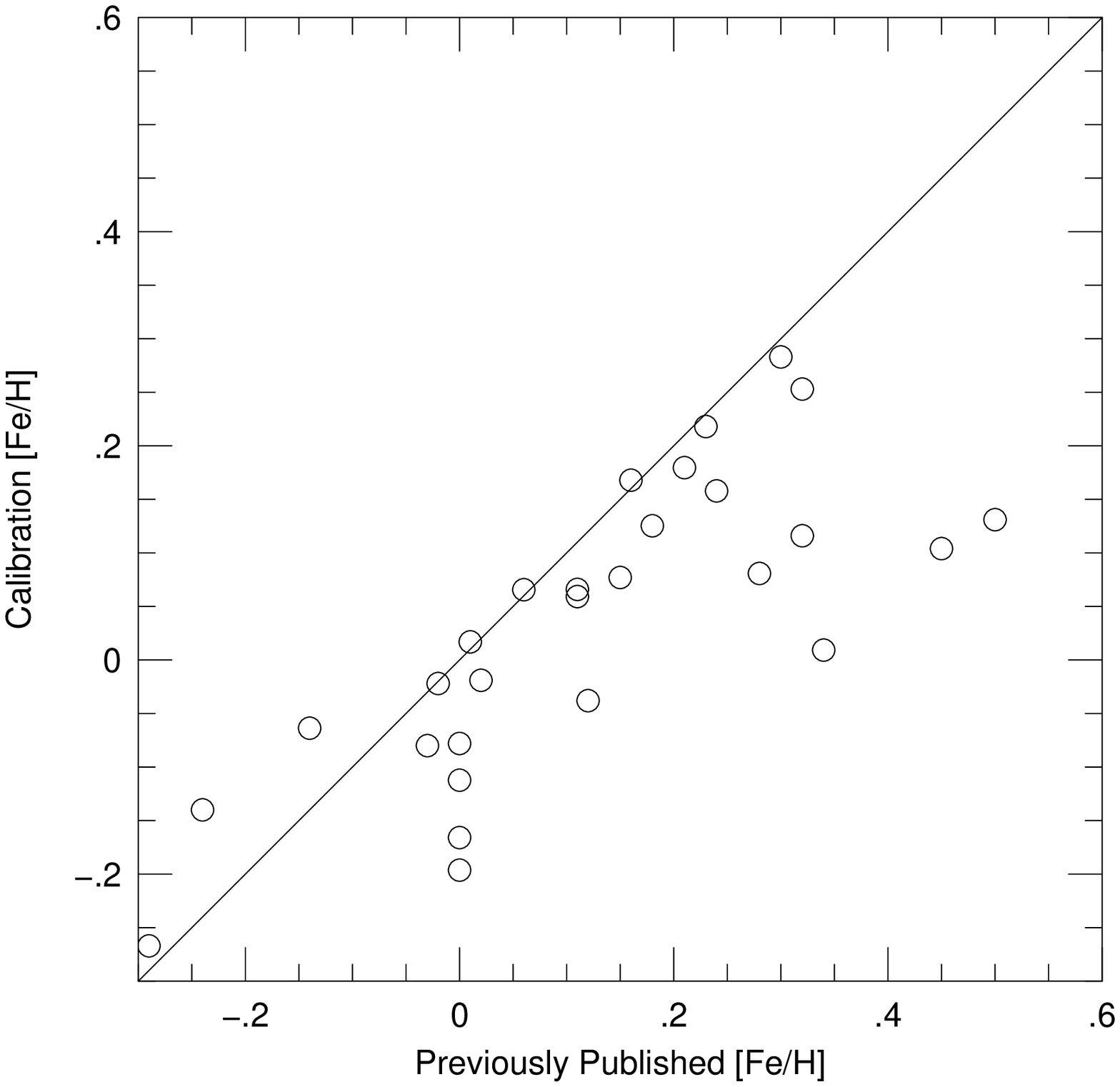}
\caption{
$uvby$ metallicities for planet-bearing stars
obtained with the Schuster \& Nissen (1989) calibration
in comparison with previously determined metallicities
compiled by Butler et al. 2000.}
\end{figure}
\clearpage

\begin{figure}
\plotone{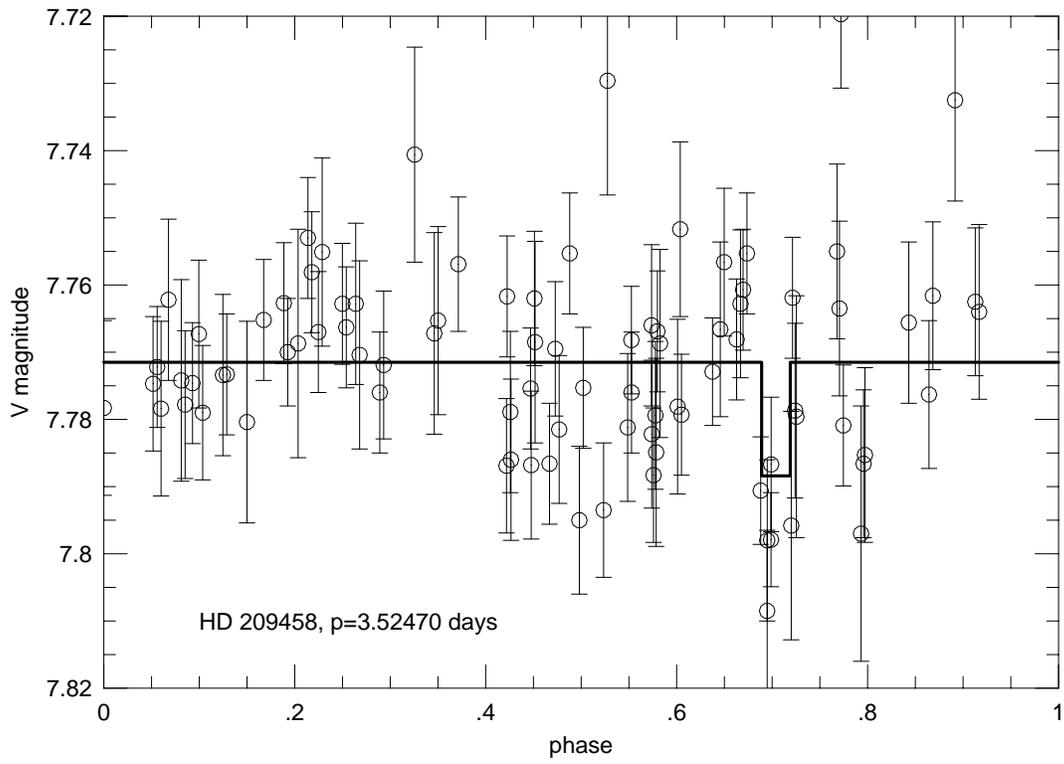}
\caption{
89 measurements from the
Hipparcos epoch photometry for HD 209458, folded at a period of 3.5247
days. An idealized transit lightcurve for the planet is shown. 
}
\end{figure}
\clearpage

\begin{figure}
\plotone{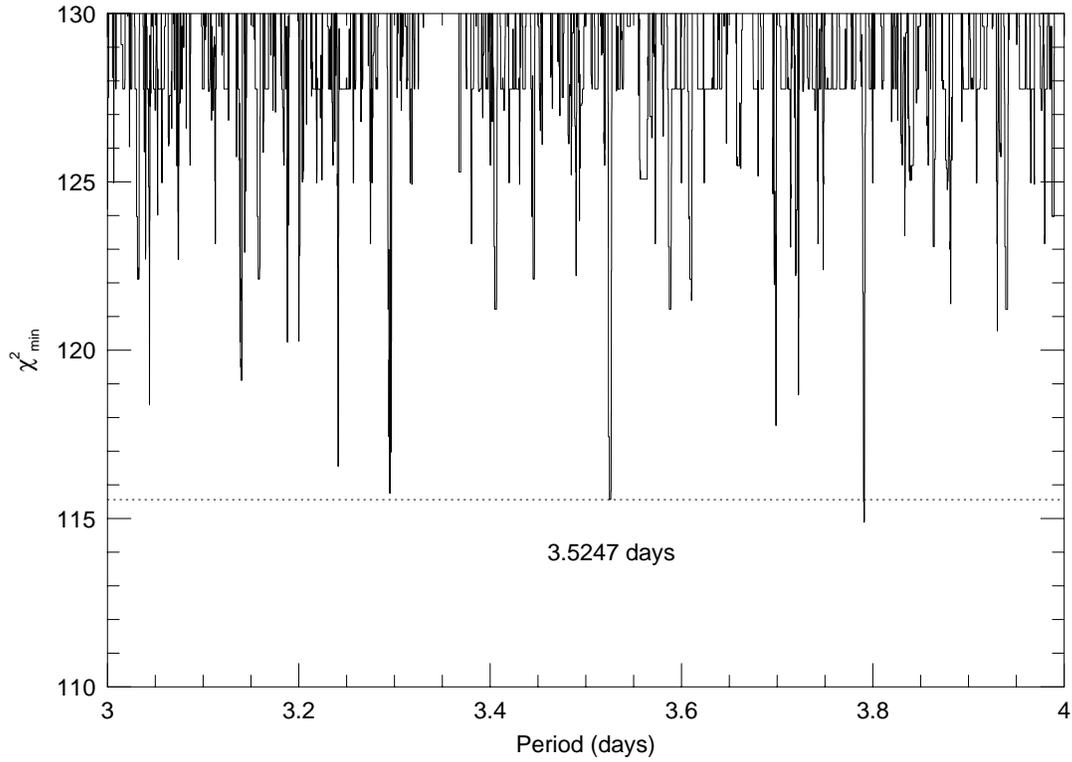}
\caption{
Minimum $\chi^{2}$ fits to the Hipparcos epoch photometry for HD 209458.
The true detection at period $P$=3.5247 days is indicated, and the
level of the corresponding minima is indicated by the dashed line. Note that
a spurious transit interval of $P$=3.7909 days has a slightly lower $\chi^{2}$ 
fit to the aggregate of photometric measurments.
}
\end{figure}
\clearpage

\begin{figure}
\plotone{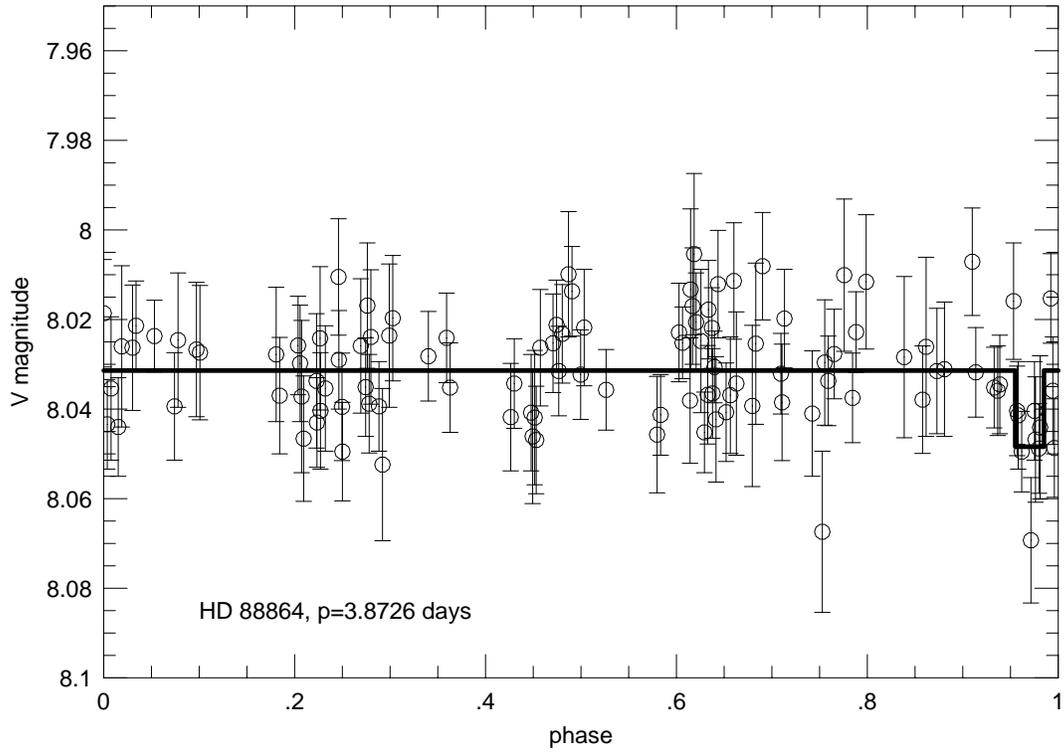}
\caption{
Hipparcos epoch photometry for HD 88864, folded at a period of 3.8726
days. There are 114 photometric observations. The lightcurve of a transit
candidate event is shown. HD 88864 is an F8V star with a $uvby$ metallicity
[{\rm Fe/H}]=0.235.
}
\end{figure}
\clearpage

\begin{figure}
\plotone{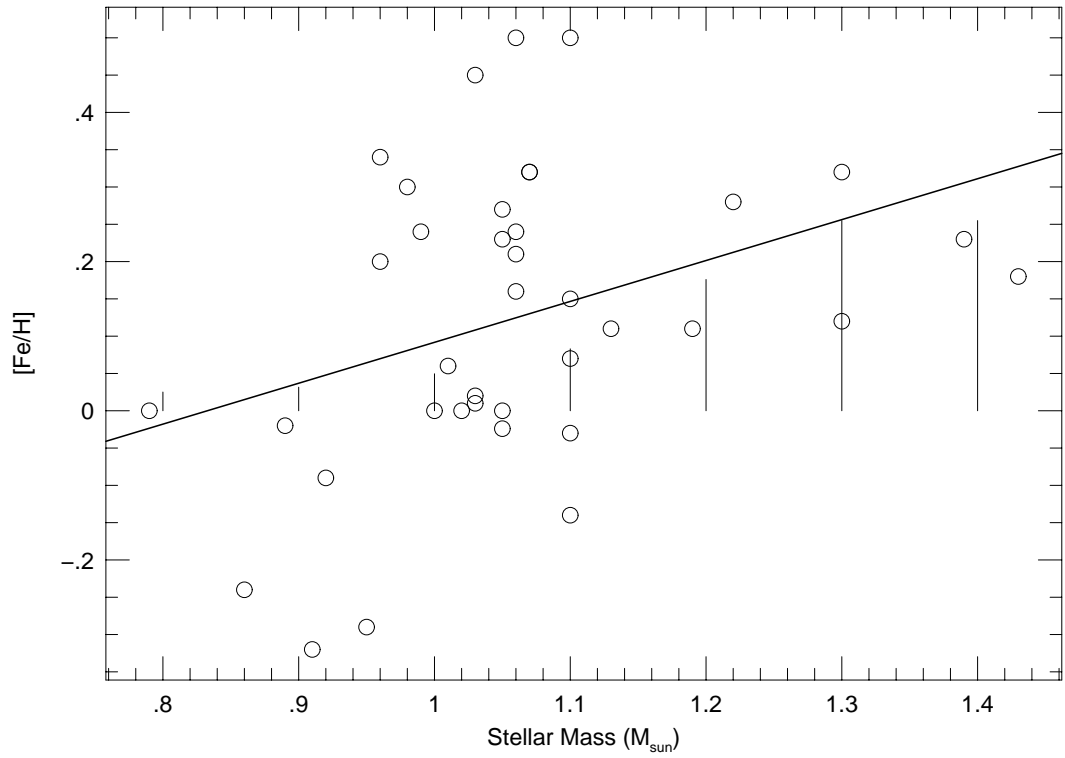}
\caption{
Metallicity vs. Stellar Mass for the known planet-bearing stars. A best-fit
slope to the data is shown, illustrating a possible trend toward
increasing metallicity as a function of stellar mass among the stars
plotted. The increase in metallicity achieved by adding 30 $M_{\oplus}$
of heavy elements to the convective envelopes of various stars
of solar metallicity are shown as vertical lines whose height corresponds
to the metallicity increase.
The 30 $M_{\oplus}$ of debris added to the fully radiative 1.4 $M_{\odot}$ star
is assumed to mix with $0.004 M_{\odot}$ of the stellar envelope.
}
\end{figure}
\clearpage

\begin{figure}
\plotone{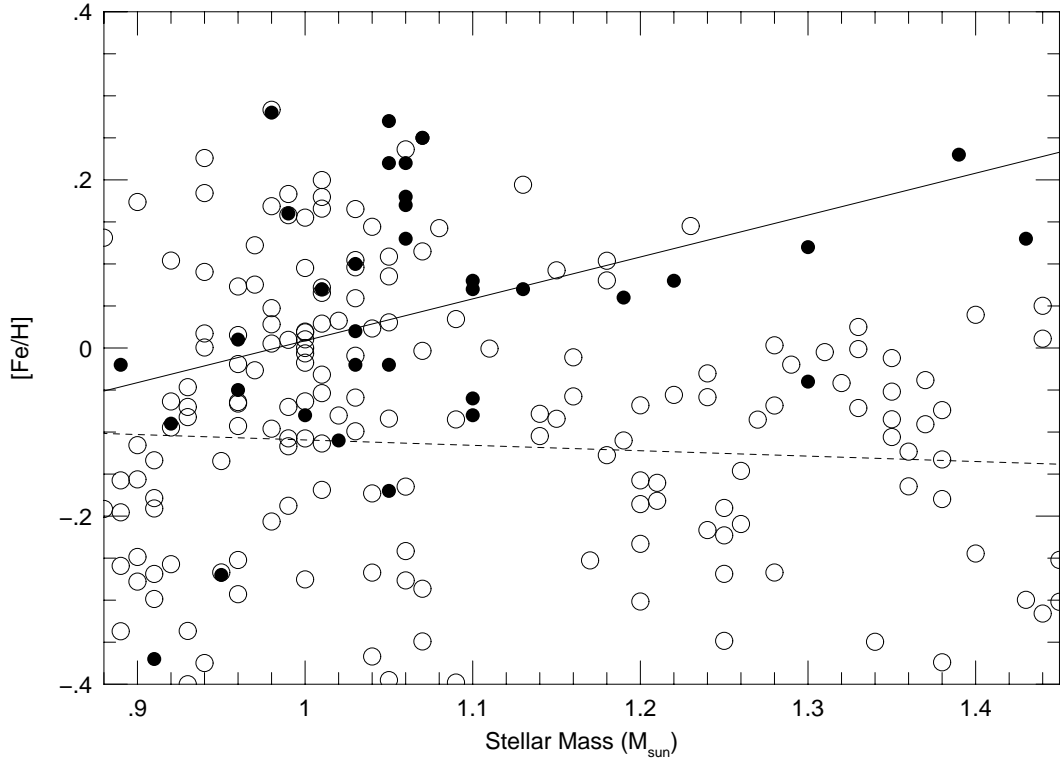}
\caption{
Metallicity vs. Stellar Mass for a volume-limited sample of nearby
stars with $M>0.88 M_{\odot}$ and metallicity $[{\rm Fe/H}] \ge -0.4$
(open circles). A best-fit slope through the data is shown as a
dashed line. Also shown are the $uvby$ metallicities of the known
planet-bearing stars (filled circles). A best-fit slope to this data
is shown as a solid line.
}
\end{figure}
\clearpage

\begin{figure}
\plotone{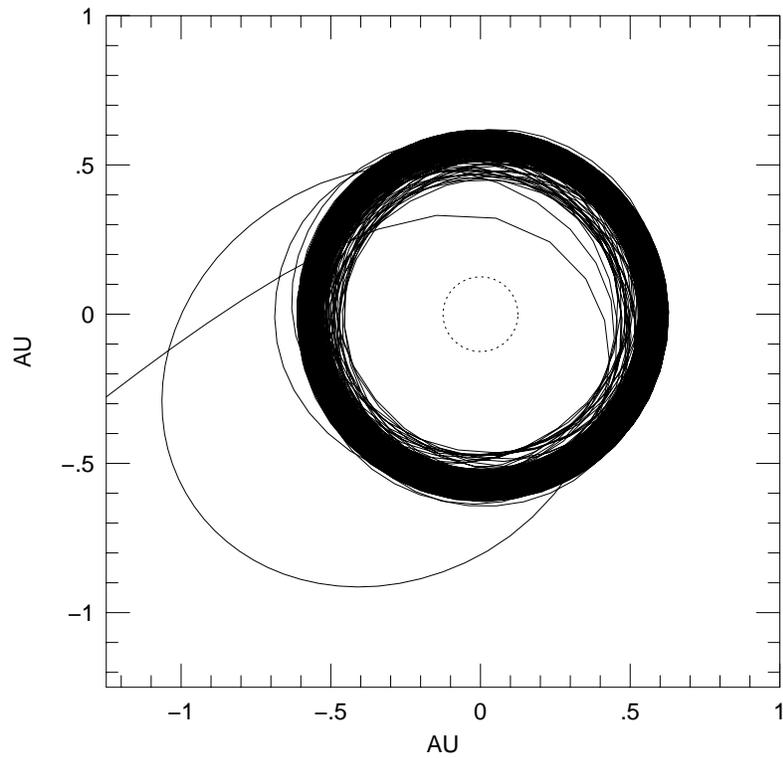}
\caption{
Illustration of a three-body encounter leading to ejection of a 
planetesimal from a binary system. The orbit of the equal mass, $e=0.$
central binary is shown as the dashed central circle. The trajectory of
a test particle placed in orbit around the binary center of mass
(and subjected to a small, constant azimuthal torque)
is shown by a thin solid line.
}
\end{figure}
\clearpage

\begin{figure}
\plotone{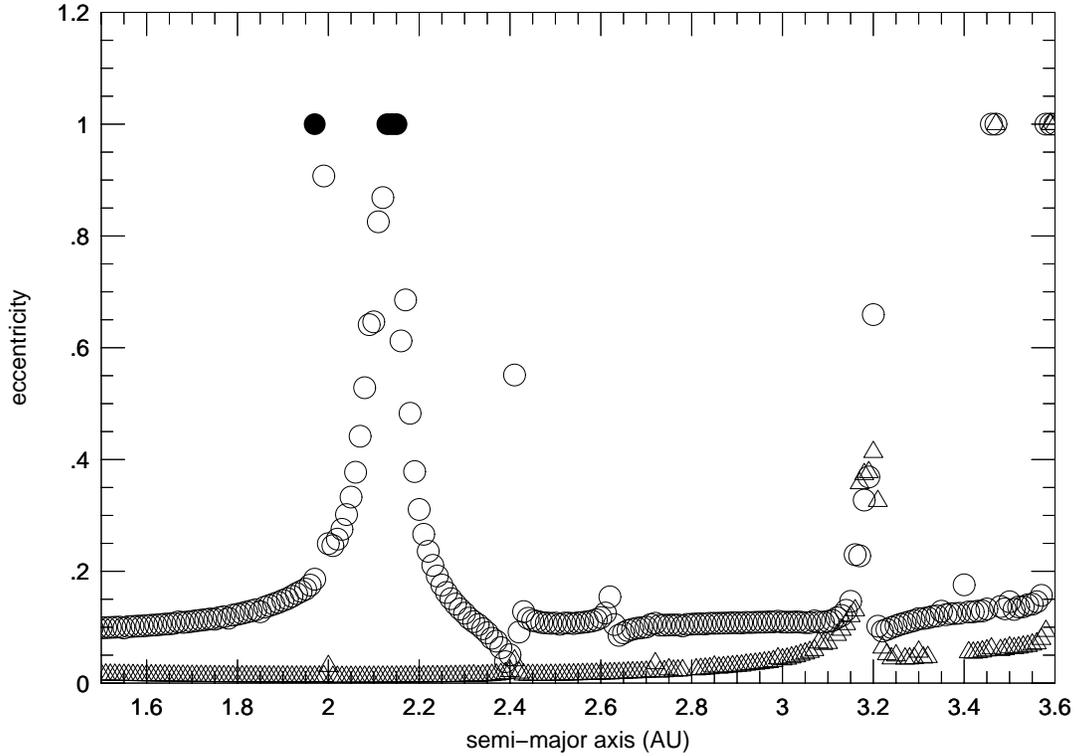}
\caption{
Maximum eccentricity of test particles placed in a model planetary
system with two planets and (a) a single central star, and (b)
a central binary with eccentricity $e$=0. and separation $a$=0.25 AU.
Circles show the maximum eccentricities
obtained for particles placed on circular orbits
with a range of initial semi-major axes. Ejections are
shown as open circles with $e$=1. Collisions with the central star
are shown as filled circles. The triangles show the maximum
eccentricities (with respect to the center of mass) of particles
placed in circumbinary orbits over a range of semi-major axes. Ejections
are shown as open triangles with $e$=1. No collisions with the 
central stars occured in the circumbinary simulations.
}
\end{figure}
\clearpage

\newpage
\centerline{\bf Table 1} 
\medskip 
\centerline{\bf Short Period Planets and their Parent Stars} 
\bigskip 
\centerline{ } 
$$
\matrix{ 
\hline
\hline
\cr
{\rm Star} & {\rm Spectra} & {\rm Period} & {\it e} & M \sin(i) & [{\rm Fe/H}]  \cr
\cr
\hline
\cr
{\rm HD} 83443   & {\rm K0 V}    & 2.986 & 0.00 & 0.30 &  --  \cr
{\rm HD} 46375   & {\rm K1 IV-V} & 3.024 & 0.00 & 0.24 &  0.34\cr
{\rm HD} 187123  & {\rm G3 V}    & 3.097 & 0.01 & 0.54 &  0.16\cr
{\rm HD} 120136  & {\rm F7 V}    & 3.313 & 0.02 & 4.14 &  0.32\cr
{\rm BD}-10 3166 & {\rm G4 V}    & 3.487 & 0.05 & 0.48 &  0.50\cr
{\rm HD} 75289   & {\rm G0 V}    & 3.508 & 0.00 & 0.46 &  0.28\cr
{\rm HD} 209458  & {\rm G0 V}    & 3.524 & 0.02 & 0.63 &  0.00\cr
{\rm HD} 217014  & {\rm G2 V}    & 4.231 & 0.01 & 0.46 &  0.21\cr
{\rm HD} 9826    & {\rm F8 V}    & 4.617 & 0.02 & 0.68 &  0.12\cr
{\rm HD} 168746  & {\rm G5}      & 6.400 & 0.00 & 0.24 & -0.09\cr
{\rm HD} 217107  & {\rm G7 V}    & 7.130 & 0.14 & 1.29 &  0.30\cr
{\rm HD} 108147  & {\rm F8/G0 V} & 10.88 & 0.56 & 0.30 & -0.02\cr
{\rm HD} 130322  & {\rm K0 V}    & 10.72 & 0.05 & 1.15 & -0.02\cr
{\rm HD} 38529   & {\rm G4 IV}   & 14.30 & 0.27 & 0.77 &  0.23\cr
{\rm HD} 75732   & {\rm G8 V}    & 14.66 & 0.03 & 0.93 &  0.45\cr
{\rm HD} 13445   & {\rm K1 V}    & 15.80 & 0.04 & 4.23 & -0.24\cr
{\rm HD} 195019  & {\rm G3 V}    & 18.20 & 0.01 & 3.55 &  0.00\cr 
\cr
\hline
}
$$

\newpage

\centerline{\bf Table 2} 
\medskip
\centerline{\bf A Catalog of Metal-Rich Stars}
$$
\matrix{
\hline
\hline
\cr
{\rm HIP \#} & {\rm HD \#} & {\rm Type} & {\rm V} & d & {\rm [{\rm Fe/H}]} & n & P & \chi^{2} & n_{t}/n_{ex} \cr
\cr
\hline
\cr
  3391&  4113&{\rm G5V     }&7.99& 44.05& .159& & & & \cr
  6978&  9070&{\rm G5      }&8.00& 42.74& .321& & & & \cr
  9035& 11833&{\rm G0      }&8.00& 87.18& .148& & & & \cr
 15005& 20155&{\rm G0V     }&7.95& 93.46& .131& & & & \cr
 19148& 25825&{\rm G0      }&7.88& 46.71& .133& & & & \cr
 20723& 28185&{\rm G5      }&7.88& 39.56& .150& & & & \cr
 30860& 45350&{\rm G5      }&7.96& 48.95& .190& & & & \cr
 32161& 48458&{\rm F3V     }&7.88& 95.06& .185& & & & \cr
 38127& 64141&{\rm F2IV    }&7.93& 96.15& .197&106& 3.5533& .939&1.29\cr
 40761& 69809&{\rm G0      }&7.93& 50.58& .140& 92& 3.0658& .971& .61\cr
 44137& 76909&{\rm G5      }&7.92& 47.66& .221& & & & \cr
 46324& 81659&{\rm G6/G8V  }&7.97& 39.89& .130& & & & \cr
 50020& 88864&{\rm F8V     }&7.97& 92.76& .235&114& 3.8726& .873&2.04\cr
 51257& 90711&{\rm K0V     }&7.94& 32.13& .174& 80& 3.1836& .861&1.96\cr
 51258& 90722&{\rm G5/G6IV }&7.96& 50.74& .271& & & & \cr
 51579& 91204&{\rm G0      }&7.89& 51.76& .171& 89& 3.1787& .896&2.26\cr
 52939& 93849&{\rm G0/G1V  }&7.92& 74.07& .240&154& 3.5060& .909&1.79\cr
 53537& 94835&{\rm G0      }&7.99& 49.46& .151& & & & \cr
 55464& 98727&{\rm F7V     }&7.91& 68.17& .187& 65& 3.7355& .998&1.31\cr
 57345&102165&{\rm F7V     }&7.88& 85.40& .360&102& 3.6098& .921&1.36\cr
\cr
\hline
}
$$
\newpage
$$
\matrix{
{\rm HIP \#} & {\rm HD \#} & {\rm Type} & {\rm V} & d & {\rm [{\rm Fe/H}]} & n & P & \chi^{2} & n_{t}/n_{ex} \cr
\cr
\hline
\cr
 59572&106156&{\rm G8V     }&8.00& 30.96& .149& & & & \cr
 60753&108351&{\rm F7V     }&7.96& 88.11& .306&116& 3.7653& .852& .74\cr
 61044&108942&{\rm G5      }&7.99& 47.57& .156& & & & \cr
 66749&118984&{\rm F3V     }&7.93& 93.28& .400&100& 3.7593& .877&2.28\cr
 81347&149724&{\rm G5      }&7.92& 55.77& .178& 94& 3.2479& .920&2.17\cr
 81767&150437&{\rm G5V     }&7.92& 55.13& .187& 50& 3.6016& .830&2.50\cr
 85017&157172&{\rm G8V     }&7.94& 33.48& .143& & & & \cr
 97336&187123&{\rm G5      }&7.92& 47.92& .168& & & & \cr
104367&201203&{\rm F8/G0V  }&7.95& 91.32& .183& 57& 3.5761& .908&1.45\cr
109836&211080&{\rm G0      }&7.89& 81.37& .162& 56& 3.1948& .860&2.06\cr
111486&213401&{\rm G5IV/V  }&7.97& 67.52& .134& & & & \cr
112187&215192&{\rm F2      }&7.97& 97.47& .167& 92& 3.5231& .971&1.78\cr
  5529&  7199&{\rm K0IV/V  }&8.12& 35.88& .126& & & & \cr
 10492& 13945&{\rm G6IV    }&8.15& 43.37& .139& & & & \cr
 10599& 13997&{\rm G5      }&8.06& 34.07& .141& 68& 3.2407& .896&2.14\cr
 16405& 21774&{\rm G5      }&8.15& 49.90& .163& & & & \cr
 28395& 40590&{\rm F6V     }&8.15& 85.91& .148& 80& 3.3677& .886&1.65\cr
 31895& 48265&{\rm G5IV/V  }&8.14& 87.41& .175& 98& 3.2525& .748&1.19\cr
 32916& 49674&{\rm G0      }&8.18& 40.73& .225& & & & \cr
 36993& 60521&{\rm G0      }&8.06& 51.92& .129& & & & \cr
 38104& 62923&{\rm G5      }&8.13& 51.47& .159& & & & \cr
 40411& 68659&{\rm F2      }&8.05& 96.25& .200& & & & \cr
 41022& 69960&{\rm G5      }&8.08& 61.20& .130& & & & \cr
 42214& 73256&{\rm G8/K0V  }&8.15& 36.52& .127& & & & \cr
 45406& 79498&{\rm G5      }&8.11& 48.64& .199& 78& 3.9984& .824&2.14\cr
\cr
\hline
}
$$
\newpage
$$
\matrix{
{\rm HIP \#} & {\rm HD \#} & {\rm Type} & {\rm V} & d & {\rm [{\rm Fe/H}]} & n & P & \chi^{2} & n_{t}/n_{ex} \cr
\cr
\hline
\cr
 49060& 86680&{\rm G0V     }&8.05& 99.70& .262& & & & \cr
 58813&104760&{\rm G2/G3III}&8.09& 55.68& .262& & & & \cr
 59382&105844&{\rm G5      }&8.14& 42.90& .147& & & & \cr
 60081&107148&{\rm G5      }&8.09& 51.26& .253& & & & \cr
 62583&111431&{\rm G3V     }&8.09& 83.61& .227& 86& 3.5510& .839&1.44\cr
 81421&149933&{\rm G5      }&8.12& 37.65& .143& & & & \cr
 91332&171918&{\rm G0      }&8.06& 57.64& .277& & & & \cr
 94625&179699&{\rm F8/G0V  }&8.08& 79.94& .141& & & & \cr
 94718&180556&{\rm F8      }&8.16& 76.10& .196&121& 3.0382& .937&1.02\cr
 99727&192343&{\rm G4V     }&8.09& 65.83& .773& 81& 3.1428& .973&2.46\cr
103692&200078&{\rm G5      }&8.13& 52.69& .157& & & & \cr
104903&202206&{\rm G6V     }&8.15& 46.34& .236&137& 3.6257& .987& .92\cr
105906&205521&{\rm G5      }&8.13& 46.99& .199& & & & \cr
106006&204313&{\rm G5V     }&8.09& 47.30& .130& & & & \cr
109169&211681&{\rm G5      }&8.17& 70.87& .144&127& 3.0527& .921&1.54\cr
  3488&  4333&{\rm F5V     }&8.17& 81.23& .243& & & & \cr
  5189&  6558&{\rm F8      }&8.27& 73.75& .137& 88& 3.7859&1.000&1.62\cr
  6712&  8765&{\rm G5      }&8.22& 75.13& .196& 88& 3.0972& .813&2.25\cr
 14180& 19493&{\rm G3IV/V  }&8.24& 76.45& .129& & & & \cr
 16107& 21313&{\rm G0      }&8.24& 71.23& .252& & & & \cr
 22429& 30339&{\rm F8      }&8.27& 73.48& .238& 85& 3.7961& .913&2.02\cr
 24176& 33822&{\rm G5V     }&8.19& 55.49& .138&115& 3.1219& .844&1.98\cr
 25436& 35996&{\rm F3/F5IV }&8.23& 81.83& .256& & & & \cr
 36310& 59062&{\rm G5      }&8.20& 46.88& .147& & & & \cr
 39417& 66428&{\rm G5      }&8.33& 55.04& .201& 86& 3.0396& .786&2.27\cr
\cr
\hline
}
$$
\newpage
$$
\matrix{
{\rm HIP \#} & {\rm HD \#} & {\rm Type} & {\rm V} & d & {\rm [{\rm Fe/H}]} & n & P & \chi^{2} & n_{t}/n_{ex} \cr
\cr
\hline
\cr
 40687& 68988&{\rm G0      }&8.27& 58.82& .364&145& 3.2838& .945&1.01\cr
 41777& 72579&{\rm K0V     }&8.28& 43.18& .136& & & & \cr
 43686& 76700&{\rm G8V     }&8.23& 59.70& .138&109& 3.2804& .876&1.35\cr
 44823& 78277&{\rm G2IV    }&8.25& 91.91& .156& & & & \cr
 45841& 80685&{\rm F5/F6V  }&8.20& 77.70& .209& 80& 3.7714& .777&2.14\cr
 53301& 94482&{\rm G1/G2V  }&8.18& 77.22& .164&123& 3.0261& .975&1.36\cr
 53424& 94690&{\rm G5      }&8.31& 52.08& .128& & & & \cr
 55076& 97854&{\rm G0      }&8.29& 81.23& .133& & & & \cr
 57468&102361&{\rm F8V     }&8.22& 84.46& .318& & & & \cr
 60052&107077&{\rm F3V     }&8.29& 92.08& .127& & & & \cr
 64028&114036&{\rm G8III   }&8.20& 52.85& .134& & & & \cr
 73408&131664&{\rm G3V     }&8.19& 57.21& .156&162& 3.5112& .978&1.89\cr
 77547&141514&{\rm F8V     }&8.21& 66.09& .168& & & & \cr
 77641&141599&{\rm G6V     }&8.28& 46.38& .181& & & & \cr
 81037&148628&{\rm F8/G0V  }&8.22& 76.80& .212& & & & \cr
 82059&151329&{\rm G0      }&8.23& 53.79& .168& & & & \cr
 85454&157798&{\rm G3V     }&8.23& 92.25& .165& & & & \cr
 89247&167081&{\rm F8      }&8.32& 49.21& .179& & & & \cr
 90593&170469&{\rm G5      }&8.28& 64.98& .264& & & & \cr
 97769&188015&{\rm G5IV    }&8.29& 52.63& .209& & & & \cr
 98228&187978&{\rm G3V     }&8.29& 84.60& .165& & & & \cr
 99034&190613&{\rm G3/G5V  }&8.20& 53.48& .193& & & & \cr
111136&213472&{\rm G5      }&8.26& 64.35& .126&130& 3.7727& .944&1.24\cr
113905&218168&{\rm G5      }&8.18& 48.80& .177& & & & \cr
  4423&  5470&{\rm G0      }&8.42& 67.80& .178& 63& 3.2387& .788&1.85\cr
\cr
\hline
}
$$
\newpage
$$
\matrix{
{\rm HIP \#} & {\rm HD \#} & {\rm Type} & {\rm V} & d & {\rm [{\rm Fe/H}]} & n & P & \chi^{2} & n_{t}/n_{ex} \cr
\cr
\hline
\cr
  6498&  8328&{\rm G5      }&8.36& 78.80& .134& & & & \cr
 12797& 17152&{\rm G8V     }&8.47& 44.09& .146& 90& 3.5859& .880&2.07\cr
 15631& 21089&{\rm G3IV    }&8.36& 68.92& .142&113& 3.0146& .871&1.23\cr
 16115& 22104&{\rm G3V     }&8.40& 64.72& .154& & & & \cr
 19024& 25682&{\rm G5      }&8.39& 45.96& .133& & & & \cr
 22953& 31827&{\rm G8IV    }&8.33& 52.49& .274& & & & \cr
 27090& 38467&{\rm G3/G5V  }&8.33& 72.57& .209& & & & \cr
 27799& 39480&{\rm G5      }&8.42& 81.90& .160& & & & \cr
 38188& 64122&{\rm F6/F7V  }&8.43& 73.10& .138& & & & \cr
 38636& 64273&{\rm G5      }&8.43& 55.43& .133& 90& 3.3756& .910&1.33\cr
 45967& 80903&{\rm G5      }&8.39& 84.89& .139& 89& 3.0949& .980&2.22\cr
 46007& 81110&{\rm G3V     }&8.37& 46.77& .180& & & & \cr
 46325& 81505&{\rm G8III   }&8.49& 86.81& .237& 96& 3.3513& .891&2.17\cr
 50839& 90028&{\rm G3V     }&8.38& 83.61& .291& & & & \cr
 61880&110314&{\rm G2V     }&8.35& 69.54& .187& & & & \cr
 65747&117243&{\rm G5III   }&8.41& 66.23& .284&121& 3.6613& .977& .78\cr
 79296&145331&{\rm K0/K1III}&8.45& 65.10& .133& & & & \cr
 82757&152776&{\rm G5      }&8.46& 86.21& .140&169& 3.6598& .879&1.49\cr
106336&204807&{\rm G6IV    }&8.37& 64.68& .172& & & & \cr
107397&206683&{\rm G3IV    }&8.39& 64.68& .135& & & & \cr
112336&215460&{\rm F0      }&8.42& 92.51& .229& 78& 3.0005& .863&1.26\cr
117526&223498&{\rm G7V     }&8.40& 48.66& .140&130& 3.3603& .953&2.06\cr
  2282&  2587&{\rm G6V     }&8.53& 48.54& .172& & & & \cr
  6197&  8038&{\rm G5V     }&8.48& 52.97& .200& & & & \cr
  7221&  9331&{\rm G5      }&8.49& 51.07& .251& & & & \cr
\cr
\hline
}
$$
\newpage
$$
\matrix{
{\rm HIP \#} & {\rm HD \#} & {\rm Type} & {\rm V} & d & {\rm [{\rm Fe/H}]} & n & P & \chi^{2} & n_{t}/n_{ex} \cr
\cr
\hline
\cr
 16727& 22282&{\rm G5      }&8.59& 50.48& .160& & & & \cr
 17269& 23398&{\rm G5IV/V  }&8.52& 75.19& .214& & & & \cr
 21850& 30177&{\rm G8V     }&8.49& 54.70& .196& & & & \cr
 24660& 34386&{\rm G5III   }&8.60& 53.94& .215& & & & \cr
 30377& 45133&{\rm G5V     }&8.47& 64.81& .156&114& 3.9419& .972&2.33\cr
 37309& 61686&{\rm G3V     }&8.61& 72.20& .166& & & & \cr
 46871& 82606&{\rm G0      }&8.59& 67.34& .170& 84& 3.7138& .920&2.28\cr
 48143& 85249&{\rm F7V     }&8.58& 88.42& .203&137& 3.5194& .957&2.39\cr
 51877& 91702&{\rm G5      }&8.54& 52.06& .185& & & & \cr
 53624& 94861&{\rm G0      }&8.51& 90.33& .136& & & & \cr
 59968&106937&{\rm G6/G8V  }&8.56& 78.43& .149& & & & \cr
 60096&107181&{\rm G3V     }&8.49& 82.51& .170& 96& 3.0243& .837&1.16\cr
 64955&115762&{\rm G2V     }&8.59& 60.35& .129& & & & \cr
 70435&126530&{\rm G0      }&8.54& 74.91& .169&113& 3.7472& .913&2.27\cr
 71103&127423&{\rm G0V     }&8.61& 69.06& .196&106& 3.4208& .906&2.28\cr
 80936&149028&{\rm G       }&8.57& 48.50& .126& & & & \cr
 81269&149396&{\rm G5IV/V  }&8.59& 56.24& .127& 51& 3.5559& .737&2.43\cr
 87679&162907&{\rm K0V     }&8.63& 48.17& .141& & & & \cr
 89321&166745&{\rm G5V     }&8.59& 52.03& .143& 74& 3.9670& .964&1.80\cr
100363&193795&{\rm G4IV    }&8.57& 71.33& .245& & & & \cr
104399&201364&{\rm A5      }&8.41& 42.25& .470& & & & \cr
105063&202697&{\rm G5      }&8.64& 83.13& .168&185& 3.2085& .869&1.88\cr
  3502&  4203&{\rm G5      }&8.75& 77.82& .225& & & & \cr
 12198& 16275&{\rm G5      }&8.72& 74.46& .188& & & & \cr
 17054& 23127&{\rm G2V     }&8.64& 89.13& .259&107& 3.9807& .850&2.18\cr
\cr
\hline
}
$$
\newpage
$$
\matrix{
{\rm HIP \#} & {\rm HD \#} & {\rm Type} & {\rm V} & d & {\rm [{\rm Fe/H}]} & n & P & \chi^{2} & n_{t}/n_{ex} \cr
\cr
\hline
\cr
 24110& 33811&{\rm G8IV/V  }&8.79& 55.99& .250& & & & \cr
 43618& 75880&{\rm G0      }&8.69& 60.64& .125& & & & \cr
 49387& 87000&{\rm G5      }&8.80& 40.50& .143& & & & \cr
 51664& 91348&{\rm G8III   }&8.73& 89.61& .252& & & & \cr
 58656&104437&{\rm G5IV    }&8.71& 64.72& .182&186& 3.5137& .991&1.32\cr
 59278&105618&{\rm G0      }&8.71& 66.76& .259&106& 3.7422& .987&1.81\cr
 62350&111069&{\rm G5      }&8.74& 59.49& .165& & & & \cr
 72041&129401&{\rm G8      }&8.76& 92.68& .260& 99& 3.7066& .894&2.25\cr
 81369&149194&{\rm F8V     }&8.72& 97.66& .386&114& 3.4416& .989&1.18\cr
104182&201093&{\rm G5      }&8.69& 96.90& .175& & & & \cr
109355&210312&{\rm G5      }&8.71& 55.90& .179& & & & \cr
116517&221954&{\rm G5/G6V  }&8.82& 98.81& .211& 96& 3.1993& .940&1.50\cr
116984&222697&{\rm G5      }&8.73& 43.76& .143& & & & \cr
  5807&  7487&{\rm F7V     }&8.86& 94.43& .168& & & & \cr
  5881&  7483&{\rm G5      }&8.94& 59.17& .254& 94& 3.7366& .909&2.04\cr
 19126& 26071&{\rm G5IV/V  }&8.97& 91.16& .130&137& 3.9492& .964&1.70\cr
 21889& 30295&{\rm K0/K1V  }&8.94& 58.34& .150& & & & \cr
 23069& 31609&{\rm G5      }&8.99& 50.86& .158& 90& 3.0239& .924&1.85\cr
 54104& 96020&{\rm G8V     }&8.89& 77.34& .278& & & & \cr
 54381& 96529&{\rm G5      }&8.93& 77.10& .158& & & & \cr
 61028&108874&{\rm G5      }&8.84& 68.54& .138& 96& 3.2223& .992&2.11\cr
 65882&117497&{\rm G0III   }&8.80& 78.62& .167& & & & \cr
 66621&118914&{\rm G0      }&8.90& 73.75& .304&143& 3.1064& .961&1.78\cr
 69660&124595&{\rm G1/G2V  }&8.85& 76.86& .148& 68& 3.1879& .683&2.12\cr
 72203&129445&{\rm G6V     }&8.85& 67.61& .246& & & & \cr
\cr
\hline
}
$$
\newpage
$$
\matrix{
{\rm HIP \#} & {\rm HD \#} & {\rm Type} & {\rm V} & d & {\rm [{\rm Fe/H}]} & n & P & \chi^{2} & n_{t}/n_{ex} \cr
\cr
\hline
\cr
109207&209913&{\rm G8V     }&8.86& 69.64& .158& & & & \cr
109428&210392&{\rm G0      }&8.80& 90.50& .280& & & & \cr
116756&222342&{\rm F0      }&8.90& 79.05& .191& & & & \cr
117427&223315&{\rm G5V     }&8.87& 53.62& .203& & & & \cr
  4752&  5946&{\rm G5      }&9.01& 91.24& .154& 78& 3.2105& .875&1.85\cr
  5389&  6880&{\rm G8/K0V  }&9.20& 46.36& .248& & & & \cr
 20273& 27496&{\rm G5      }&9.03& 92.51& .133& & & & \cr
 21703& 29528&{\rm K0      }&8.99& 56.69& .206& & & & \cr
 27688& 39503&{\rm G5IV/V  }&9.04& 63.53& .138& & & & \cr
 29550& 43197&{\rm G8/K0IV }&9.03& 54.95& .180&123& 3.6620& .995&2.05\cr
 42084& 72680&{\rm K0      }&9.00& 51.63& .200& 64& 3.0322& .737&2.37\cr
 44279& 77417&{\rm G8IV    }&9.10& 65.06& .192& & & & \cr
 45982& 80606&{\rm G5      }&9.01& 58.38& .195& & & & \cr
 46113& 81270&{\rm F6V     }&9.01& 91.83& .253& 88& 3.2640& .925&2.36\cr
 62198&110855&{\rm G0      }&9.11& 80.65& .138&101& 3.3616& .883&1.77\cr
101982&340795&{\rm G5      }&8.99& 75.82& .182& & & & \cr
114967&219556&{\rm K0V     }&9.19& 58.17& .136& & & & \cr
  9550& 12585&{\rm G3V     }&9.29& 77.40& .233& & & & \cr
 22320& 30669&{\rm G8/K0V  }&9.17& 54.91& .142&151& 3.8691& .955&1.52\cr
 46639&233641&{\rm G0      }&9.28& 96.62& .335& & & & \cr
 57735&102843&{\rm K0      }&9.32& 54.44& .172& & & & \cr
 58318&103829&{\rm F8      }&9.29& 89.21& .274& & & & \cr
 61123&108953&{\rm G8/K0IV/}&9.36& 63.73& .176& & & & \cr
 62093&110605&{\rm G8V     }&9.21& 71.68& .162& & & & \cr
 78521&143361&{\rm G6V     }&9.24& 59.35& .204& 96& 3.2201& .789&1.81\cr
\cr
\hline
}
$$
\newpage
$$
\matrix{
{\rm HIP \#} & {\rm HD \#} & {\rm Type} & {\rm V} & d & {\rm [{\rm Fe/H}]} & n & P & \chi^{2} & n_{t}/n_{ex} \cr
\cr
\hline
\cr
 82632&152079&{\rm G6V     }&9.27& 85.18& .186& 85& 3.4298& .804&1.78\cr
 88631&165204&{\rm G6/G8V  }&9.20& 75.87& .165& & & & \cr
 97125&186265&{\rm G8V     }&9.29& 86.81& .138& 86& 3.1557& .880&2.00\cr
117882&224040&{\rm F3/F5V  }&9.27& 84.18& .354& & & & \cr
  5311&  6790&{\rm G0V     }&9.43& 87.80& .363& & & & \cr
 27113& 38554&{\rm G6V     }&9.46& 82.78& .140& & & & \cr
 48809& 86397&{\rm G8V     }&9.62& 69.40& .179& & & & \cr
 94552&179640&{\rm G6V     }&9.56& 91.16& .201& & & & \cr
 94615&230999&{\rm G5      }&9.76& 99.30& .410& & & & \cr
104226&200869&{\rm K1IV/V  }&9.48& 63.65& .206& & & & \cr
 33634& 52217&{\rm G8V     }&9.98& 74.35& .156& & & & \cr
\cr
\hline
}
$$

{}
\end{document}